\pdfoutput=1
\documentclass[usenatbib]{mn2e}
\usepackage{siunitx}
\usepackage{graphicx}
\usepackage{latexsym}
\usepackage{aas_macros}
\usepackage{amsfonts,amsmath,amssymb}
\usepackage{url}
\usepackage[utf8]{inputenc}
\usepackage{fancyref}

\title[PHL 1445: An eclipsing CV near the period minimum]{PHL 1445: An eclipsing cataclysmic variable with a substellar donor near the period minimum}

\author[M.\,J.\ McAllister et al.]{M.\,J.\ McAllister$^{1}$, S.\,P.\ Littlefair$^{1}$, I.\ Baraffe$^{2}$, V.\,S.\ Dhillon$^{1}$, T.\,R.\ Marsh$^{3}$, 
\newauthor J.\ Bento$^{7}$, J.\ Bochinski$^{6}$, M.\,C.\,P.\ Bours$^{3}$, E.\ Breedt$^{3}$, C.\,M.\ Copperwheat$^{4}$,
\newauthor L.\,K.\ Hardy$^{1}$, P.\ Kerry$^{1}$, S.\,G.\ Parsons$^{5}$, J.\,W.\ Rostron$^{3}$, D.\,I.\ Sahman$^{1}$, 
\newauthor C.\,D.\,J.\ Savoury$^{1}$, R.\,L.\ Tunnicliffe$^{3}$\\
$^{1}$Dept of Physics and Astronomy, University of Sheffield, Sheffield, S3 7RH, UK\\
$^{2}$Dept of Physics and Astronomy, University of Exeter, Exeter, EX4 4QL, UK\\
$^{3}$Dept of Physics, University of Warwick, Coventry, CV4 7AL, UK\\
$^{4}$Astrophysics Research Institute, Liverpool John Moores University, Liverpool, L3 5RF, UK\\
$^{5}$Departmento de F\'isica y Astronom\'ia, Universidad de Valpara\'iso, Avenida Gran Bretana 1111, Valpara\'iso, 2360102, Chile\\
$^{6}$Department of Physical Sciences, The Open University, Milton Keynes, MK7 6AA, UK\\
$^{7}$Department of Physics and Astronomy, Macquarie University, NSW 2109, Australia}

\begin{document}

%\date{Accepted 1988 December 15. Received 1988 December 14; in original form 1988 October 11}

%\pagerange{\pageref{firstpage}--\pageref{lastpage}} \pubyear{2002}

\maketitle

\label{firstpage}

\begin{abstract}

We present high-speed, three-colour photometry of the eclipsing dwarf nova PHL 1445, which, with an orbital period of 76.3\,min, lies just below the period minimum of $\sim$82\,min for cataclysmic variable stars. Averaging four eclipses reveals resolved eclipses of the white dwarf and bright spot. We determined the system parameters by fitting a parameterised eclipse model to the averaged lightcurve. We obtain a mass ratio of $q$=$0.087\pm0.006$ and inclination $i$=$85.2\pm0.9^{\circ}$. The primary and donor masses were found to be $M_{w}$=$0.73\pm0.03\,M_{\odot}$ and $M_{d}$=$0.064\pm0.005\,M_{\odot}$, respectively. Through multicolour photometry a temperature of the white dwarf of $T_{w}$=$13200\pm700$\,K and a distance of $220\pm50$\,pc were determined.

The evolutionary state of PHL 1445 is uncertain. We are able to rule out a significantly evolved donor, but not one that is slightly evolved. Formation with a brown dwarf donor is plausible; though the brown dwarf would need to be no older than 600\,Myrs at the start of mass transfer, requiring an extremely low mass ratio ($q$=0.025) progenitor system. PHL 1445 joins SDSS 1433 as a sub-period minimum CV with a substellar donor. These existence of two such systems raises an alternative possibility; that current estimates for the intrinsic scatter and/or position of the period minimum may be in error.

\end{abstract}


\bibliographystyle{mn2e_fixed2}

\section{Introduction} 
\label{sec:introduction}

Cataclysmic variable stars (CVs) are close binary systems, with each system containing a white dwarf primary and low mass secondary. The secondary star is large enough to fill its Roche lobe and therefore mass is transferred to the white dwarf. In systems with a low magnetic-field white dwarf, this transferred mass does not immediately accrete onto the surface of the white dwarf, instead forming an accretion disc around it in order to conserve angular momentum. A bright spot forms where the gas stream from the donor impacts the disc. For a general review of CVs, see \cite{hellier01}.

The structure of CVs can, at some inclinations, result in complex eclipses, with the accretion disc, white dwarf and bright spot all being eclipsed by the secondary star in quick succession. High-time resolution photometry allows each of these individual features to be observed and their timings determined, which can then be used to determine accurate system parameters \citep[eg.][]{wood86}.

Steady mass transfer from the donor secondary to the white dwarf primary is possible due to angular momentum loss from the system. Without angular momentum loss, mass loss from the donor increases the size of the Roche lobe until it is no longer filled by the donor, causing mass transfer to cease. Angular momentum loss reduces the size of the Roche lobe, countering the effect of donor mass loss and allowing steady mass transfer. Mass transfer in CVs leads to an evolution towards smaller system separations and therefore shorter orbital periods.

As CVs evolve to shorter orbital periods, their donors are driven further away from thermal equilibrium. This is a consequence of mass loss from the donor, more specifically a consequence of the donor's thermal time scale increasing at a more rapid rate than its mass loss time scale. As mass continues to be transferred from the donor, it eventually enters the substellar regime, and this is approximately where it is far enough away from thermal equilibrium for its radius to no longer decrease in response to further mass loss. The degenerate nature of the substellar donor can even cause its radius to increase in response to mass loss, resulting now in an increasing system separation and orbital period.

CV evolution theory therefore predicts the existence of an orbital period minimum, and this is what is observed, with the period minimum currently estimated to be at $81.8 \pm 0.9$\,min \citep{knigge11}. An accumulation of systems is also expected to be found at the period minimum - the ``period spike'' - due to systems spending more time at this stage in their evolution. This feature has been observed at $82.4 \pm 0.7$\,min \citep{gaensicke09}, in excellent agreement with the period minimum. There are, however, a handful of CVs that have periods below this period minimum.

An example of such a CV, with an orbital period of 76.3\,min, is PHL 1445. PHL 1445 was first catalogued as a faint blue object by \cite{haro62} in the Palomar-Haro-Luyten catalogue, and again (as PB 9151) by \cite{berger84}. It was identified as a CV system by \cite{wils09} through spectroscopic analysis of the 6dF Galaxy Survey target 6dFGS g0242429-114646, found to be coincident with PHL 1445. Its spectrum showed double-peaked emission lines, indicating a high inclination and possibly deeply eclipsing system \citep{wils09}. The eclipsing nature of PHL 1445 was confirmed by \cite{wils11} through follow-up photometry, which also gave the first determination of the system's orbital period.

There are multiple ways for a hydrogen rich CV to have an orbital period shorter than the period minimum. These include Galactic halo membership \citep{patterson08,uthas11}, an evolved donor \citep{thorstensen02,podsiadlowski03} or formation with a brown dwarf donor \citep{kolbbaraffe99,politano04}. Obtaining PHL 1445's donor mass and temperature may help reveal why it lies below the period minimum.

In this paper we present high-time resolution {\sc ultracam} eclipse lightcurves of PHL 1445, with system parameters determined through lightcurve modelling of an average lightcurve. Individual lightcurves are also given the same treatment, in order to see how certain parameters vary between eclipses. The observations are described in section~\ref{sec:observations}, the results displayed in section~\ref{sec:results}, and the analysis of these results in section~\ref{sec:discussion}.

\section{Observations}
\label{sec:observations}

PHL 1445 was observed over seven observing runs (Aug 2011 - Jan 2014) using {\sc ultracam} \citep{dhillon07} on the 4.2-m William Herschel Telescope (WHT), La Palma. Fifteen eclipses were observed in total, the majority observed simultaneously in the SDSS-$u'g'r'$ colour bands, the rest in SDSS-$u'g'i'$. A complete journal of observations is shown in Table~\ref{table:obs}.

Data reduction was carried out using the {\sc ultracam} pipeline reduction software (see \citealt{feline04}). A nearby, photometrically stable comparison star was used to correct for any transparency variations during observations. The standard stars Feige 22 (observed at the start of the night on 16th Jan 2012) and SA92-342 (observed at the end of the night on 30th Jul 2013) were used to transform the photometry into the $u'g'r'i'z'$ standard system \citep{smith02}.

The photometry was corrected for extinction using nightly measurements of the $r'$-band extinction from the Carlsberg Meridian Telescope\footnote{\url{http://www.ast.cam.ac.uk/ioa/research/cmt/camc_extinction.html}}, and subsequently converted into $u'$, $g'$ and $i'$-band extinction using the information provided in La Palma Technical Note 31.\footnote{\url{http://www.ing.iac.es/Astronomy/observing/manuals/ps/tech_notes/tn031.pdf}}

The typical out-of-eclipse photometric errors were estimated at 4\%, 2\% \& 2\% in the $u'$, $g'$ and $r'$ bands, respectively. These errors increased to approximately 12\%, 8\% \& 7\% when both the white dwarf and bright spot were eclipsed.

\begin{table*}
\begin{center}
\begin{tabular}{lcccccccccc}
\hline
Date & Start Phase & End Phase & Filters & $T_{mid}$ & $T_{exp}$ & NBLUE & $N_{exp}$ & Seeing & Airmass & Phot?\\
&&&& (HMJD) & (seconds) &&& (arcsecs) && \\ \hline
2011 Aug 26 & -1264.338 & -1263.098 & $u' g' i'$ & 55800.15106(3) & 2.685 & 2 & 2098 & 0.8-1.7 & 1.35-1.69 & No \\
2011 Nov 01 & -0.604 & 0.131& $u' g' r'$ & 55867.12400(3) & 2.137 & 2 & 1119 & 1.0-2.7 & 1.39-1.51 & No \\
2012 Jan 14 & 1391.856 & 1392.217 & $u' g' r'$ & 55940.87898(3) & 1.979 & 3 & 827 & 1.2-2.5 & 1.33-1.37 & No \\
2012 Jan 14 & 1393.696 & 1394.177 & $u' g' r'$ & 55940.98490(3) & 1.979 & 3 & 1102 & 1.3-3.5 & 1.90-2.36 & No \\
2012 Jan 15 & 1412.678 & 1413.119 & $u' g' r'$ & 55941.99163(3) & 1.979 & 3 & 1008 & 1.1-1.9 & 2.03-2.53 & Yes \\
2012 Jan 16 & 1429.792 & 1430.500 & $u' g' r'$ & 55942.89237(3) & 1.979 & 3 & 1619 & 1.1-6.4 & 1.36-1.80 & Yes \\
2012 Jan 16 & 1430.500 & 1431.183 & $u' g' r'$ & 55942.94534(3) & 1.979 & 3 & 1561 & 1.1-6.4 & 1.36-1.80 & Yes \\
2012 Jan 22 & 1541.725 & 1542.147 & $u' g' r'$ & 55948.82668(3) & 1.979 & 3 & 966 & 0.9-3.0 & 1.32 & Yes \\
2012 Sep 08 & 5888.678 & 5889.173 & $u' g' i'$ & 56179.15198(3) & 2.982 & 3 & 754 & 1.0-1.8 & 1.36-1.44 & No \\
2012 Oct 13 & 6546.740 & 6547.175 & $u' g' r'$ & 56214.01605(3) & 3.480 & 3 & 571 & 1.4-3.3 & 1.52-1.69 & Yes \\
2013 Jul 30 & 12023.706 & 12024.176 & $u' g' i'$ & 56504.21431(3) & 3.852 & 3 & 557 & 1.3-2.9 & 1.54-1.74 & Yes \\
2013 Dec 31 & 14925.879 & 14926.216 & $u' g' r'$ & 56657.97642(3) & 3.922 & 3 & 394 & 1.5-2.5 & 1.55-1.69 & Yes \\
2014 Jan 01 & 14941.796 & 14942.260 & $u' g' r'$ & 56658.82417(3) & 3.628 & 3 & 584 & 1.0-1.9 & 1.38-1.47 & Yes \\
2014 Jan 01 & 14944.657 & 14945.127 & $u' g' r'$ & 56658.98310(3) & 3.628 & 3 & 591 & 1.0-1.9 & 1.54-1.74 & Yes \\
2014 Jan 02 & 14960.781 & 14961.153 & $u' g' r'$ & 56659.83095(3) & 3.628 & 3 & 467 & 0.8-1.3 & 1.37-1.43 & Yes \\
\hline
\end{tabular}
\caption{\label{table:obs}Journal of observations. The dead-time between exposures was 0.025\,s for all observations. The GPS timestamp on each data point is accurate to 50\,$\mu$s. $T_{mid}$ represents the mid-eclipse time, while $T_{exp}$ \& $N_{exp}$ represent the exposure time and number of exposures, respectively. NBLUE indicates the number of $u'$-band frames which were co-added on-chip to reduce the impact of readout noise. The last column indicates whether or not the observations were photometric.}
\end{center}
\end{table*}

\section{Results}
\label{sec:results}

\subsection{Orbital ephemeris}
\label{subsec:orbeph}

Mid-eclipse times ($T_{mid}$) were determined assuming that the white dwarf eclipse is symmetric around phase zero: $T_{mid} = (T_{wi} + T_{we})/2$, where $T_{wi}$ and $T_{we}$ are the times of white dwarf mid-ingress and mid-egress, respectively. $T_{wi}$ and $T_{we}$ were determined by locating the minimum and maximum times of the smoothed $g'$-band lightcurve derivative. The $T_{mid}$ errors (see Table~\ref{table:obs}) were adjusted to give $\chi^{2}$ = 1 with respect to a linear fit. The eclipse observed on 26th Aug 2011 is the exception here, as no white dwarf features can be seen in the eclipse due to PHL 1445 being in outburst. In this instance, the time of minimum light was used for $T_{mid}$.

All eclipses, with the exception of the Aug 2011 outburst eclipse, were used to determine the following ephemeris:
\\
\\
$HMJD$ = 55867.123984(12) + 0.0529848884(13) $E$
\\
\\
This ephemeris was used to phase-fold the data for the analysis that follows.

\subsection{Lightcurve morphology and variations}
\label{subsec:lcmorph} 

\begin{figure}
\begin{center}
\includegraphics[width=1.0\columnwidth,trim=75 25 75 10]{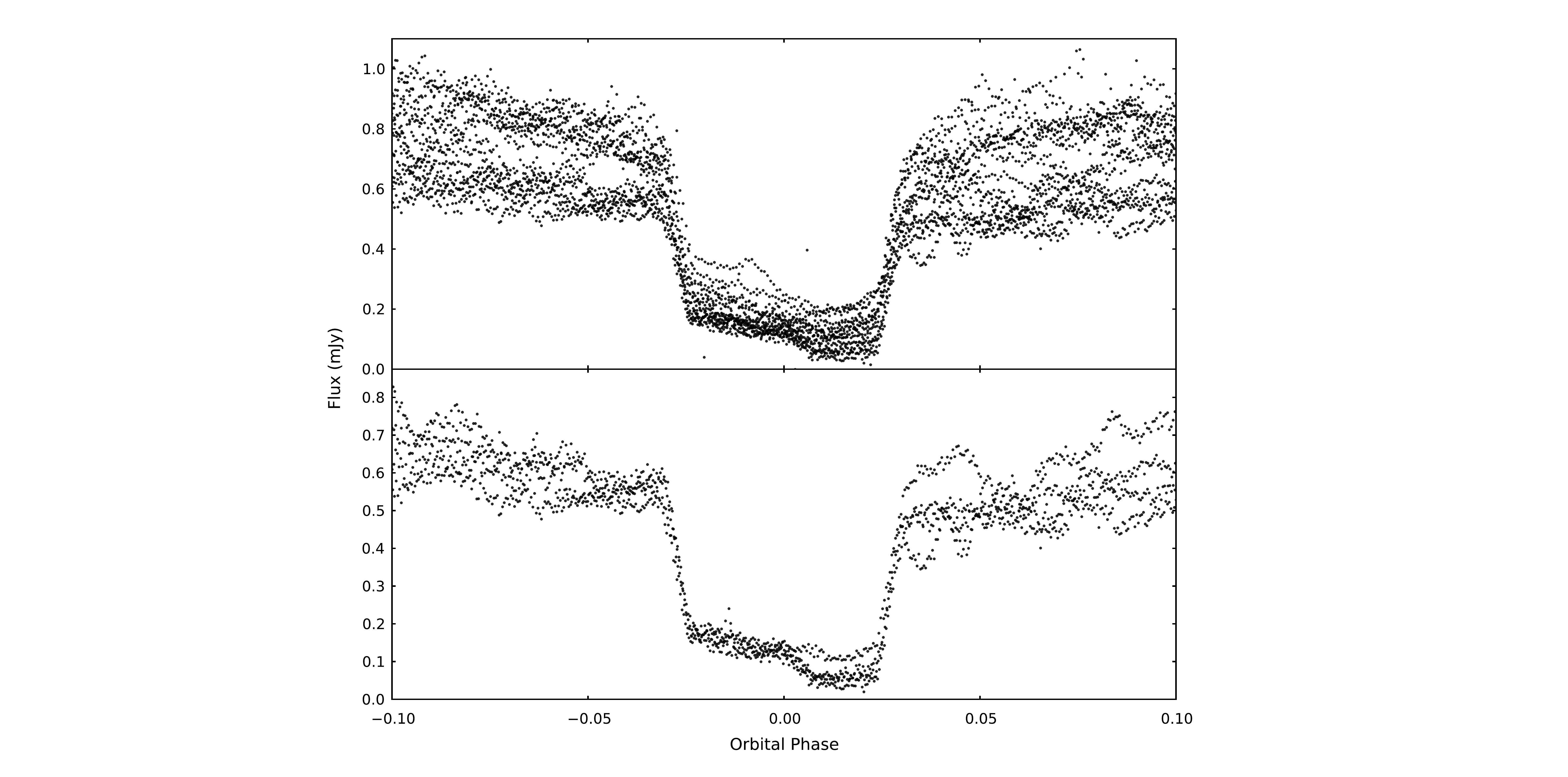}
\caption{\label{fig:pf}$Top$: All 14 quiescent PHL 1445 $g'$-band eclipses observed. $Bottom$: Four PHL 1445 $g'$-band eclipses chosen to create an average.}
\end{center}
\end{figure}

Aside from the single outburst eclipse mentioned above, all other observations listed in Table~\ref{table:obs} show a strong white dwarf eclipse feature. The same cannot be said for a bright spot feature, as a bright spot ingress can be discerned in most cases, but not one eclipse shows a clear egress. The reason for the lack of clear bright spot egress in any of these lightcurves is the strong flickering seen in this system. The flickering appears to begin immediately after white dwarf egress at around phase 0.03 (see Figure~\ref{fig:pf}), implying that its source is close to the white dwarf, either in the inner disc or boundary layer.

To help reduce the prominence of the strong flickering present, and to be able to locate the position of the bright spot egress, it was necessary to average multiple eclipses together. The 10 eclipses showing signs of a bright spot ingress were phase-folded using the ephemeris above and averaged, allowing a broad, faint bright spot egress feature to emerge. Averaging also seemed to reduce the strength of the bright spot ingress. After analysing each individual eclipse, it was apparent that the position of the bright spot ingress varied significantly across the range of observations due to changes in accretion disc radius. This is the reason for the bright spot ingress feature becoming broad and weak in the average eclipse.

To fix this issue, four eclipse lightcurves (phases 0, 1413, 1430 \& 1431) observed not too far apart in time (Nov 2011 - Jan 2012), with clear bright spot ingresses at a similar position were phase-folded and averaged. It is apparent from Figure~\ref{fig:pf} that these four chosen eclipses occurred when PHL 1445 was in the lower of two distinct photometric states (clear gap visible in top plot just before white dwarf ingress), although this was not a criterion for choosing candidates for the average. A further two eclipses were observed within this time span (both on 14th Jan), but were not used due to a mixture of bad observing conditions and lack of a visible bright spot ingress. This new average lightcurve revealed much sharper bright spot features than that consisting of all 10 eclipses, and it is worth noting that the position of the bright spot egress remained unchanged from the 10-eclipse average. The total rms of this average lightcurve's residuals is approximately 7\%, which is significantly larger than the typical photometric error ($\sim$2\%) and shows that while flickering has been decreased through averaging, it continues to be an issue. A model was then fit to this average lightcurve, in order to obtain the system parameters (Figure~\ref{fig:avefits}).

\subsection{Phase-folded average lightcurve modelling}
\label{subsec:avelcmod}

The model of the binary system used here to calculate eclipse lightcurves contains contributions from the white dwarf, bright spot, accretion disc \& secondary star, and is described in detail by \cite{savoury11}. We note briefly that the model constrains the mass ratio ($q$), white dwarf radius as a fraction of the binary separation ($R_{w}/a$), white dwarf eclipse phase full-width at half-depth ($\Delta\phi$) and white dwarf flux; these parameters can then be used to calculate system parameters (see section~\ref{subsec:syspars}). It is worth noting that we use a simplified version of the bright spot model (as described in \citealt{littlefair07}), as none of the derived system parameters were found to change significantly between models, and an F-test \citep{press07} showed the extra complexity is not justified for PHL 1445. The model requires a number of assumptions, including that of an unobscured white dwarf \citep{savoury11}. The validity of this assumption has recently been questioned by \cite{sparkodonoghue15}, through fast photometry observations of the dwarf nova OY Car. It is not yet completely clear that the results of \cite{sparkodonoghue15} cannot be explained by flickering in the boundary layer \& inner disc, and coupled with agreement between photometric \& spectroscopic parameter estimates \citep{savoury12,copperwheat12} we feel an unobscured white dwarf is still a reasonable assumption to make.

\begin{figure}
\begin{center}
\includegraphics[width=1.0\columnwidth,trim=50 25 50 25]{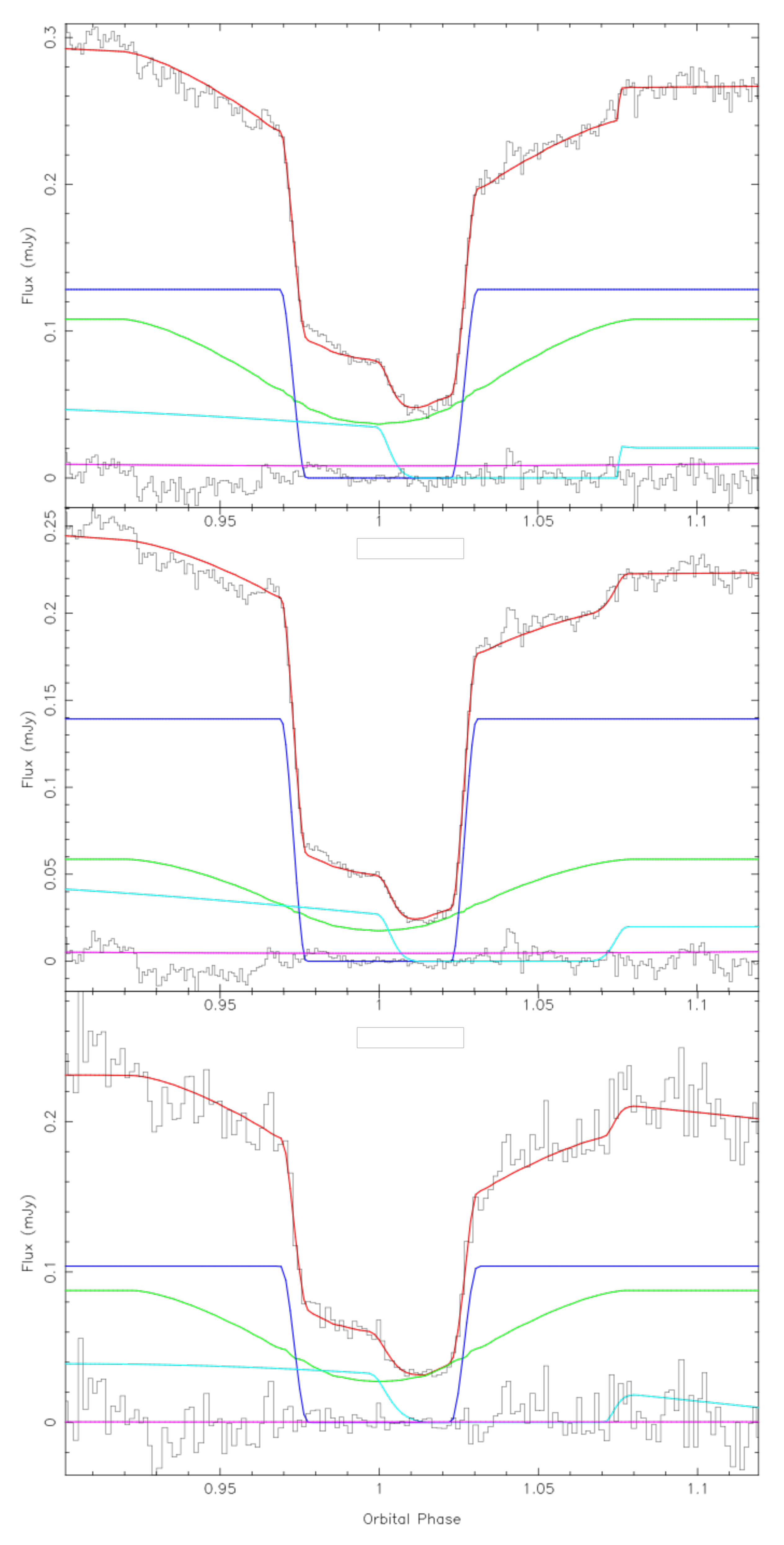}
\caption{\label{fig:avefits}Model fits (red) to average PHL 1445 lightcurves (black) in $r'$ (top), $g'$ (middle) and $u'$ (bottom) bands. Also shown are the different components of the model: white dwarf (dark blue), bright spot (light blue), accretion disc (green) and donor (purple). The residuals are at the bottom of each plot.}
\end{center}
\end{figure}

As discussed in section~\ref{subsec:lcmorph}, four PHL 1445 eclipses were phase-folded and averaged, with the resulting lightcurves in $u'$, $g'$ and $r'$ bands shown in Figure~\ref{fig:avefits}. Initial Markov chain Monte Carlo (MCMC) fits to the $u'$, $g'$ and $r'$ band data were carried out. All model parameters were left to be fit freely, apart from the white dwarf limb-darkening parameter ($U_{w}$), which was kept fixed at an initial value of 0.345. The reason for keeping $U_{w}$ fixed is because we cannot accurately constrain this parameter with the quality of data available.

The white dwarf fluxes returned from these initial model fits were then fitted - again using MCMC routines - to white dwarf atmosphere predictions \citep{bergeron95}, in order to derive initial estimates of the temperature, log\,$g$ and distance. Reddening was also included as a parameter, in order for its uncertainty to be taken into account when determining the error in temperature, but is not constrained by our data. All priors used were uninformative and uniform. Systematic errors of 5\% were added to the white dwarf fluxes returned by the model fitting, as the formal errors did not take into account any uncertainties in our absolute photometry.

With rough estimates of the white dwarf temperature and log\,$g$ known, more reliable $U_{w}$ values could be obtained using the data tables in \cite{gianninas13}. Limb-darkening parameters of 0.469, 0.390 and 0.340 were determined for $u'$, $g'$ and $r'$ bands, respectively. The typical value of 0.345 for $U_{w}$ was replaced with these new values and - again keeping $U_{w}$ fixed - the eclipse model fits were carried out for a second and final time.

\subsection{System parameters}
\label{subsec:syspars}

The mass ratio ($q$), white dwarf eclipse phase full-width at half-depth ($\Delta\phi$), and scaled white dwarf radius ($R_{w}/a$) posterior probability distributions returned by the MCMC fits described in section~\ref{subsec:avelcmod} can be used along with Kepler's third law, the system orbital period and a temperature-corrected white dwarf mass-radius relationship \citep{wood95} to calculate the posterior probability distributions of the system parameters \citep{savoury11}. These system parameters include:

\begin{enumerate}
\item mass ratio, $q$;
\item white dwarf mass, $M_{w}$;
\item white dwarf radius, $R_{w}$;
\item white dwarf log\,$g$;
\item donor mass, $M_{d}$;
\item donor radius, $R_{d}$;
\item binary separation, $a$;
\item white dwarf radial velocity, $K_{w}$;
\item donor radial velocity, $K_{d}$;
\item inclination, $i$.
\end{enumerate}

\begin{figure}
\begin{center}
\includegraphics[width=1.0\columnwidth,trim=160 80 130 40]{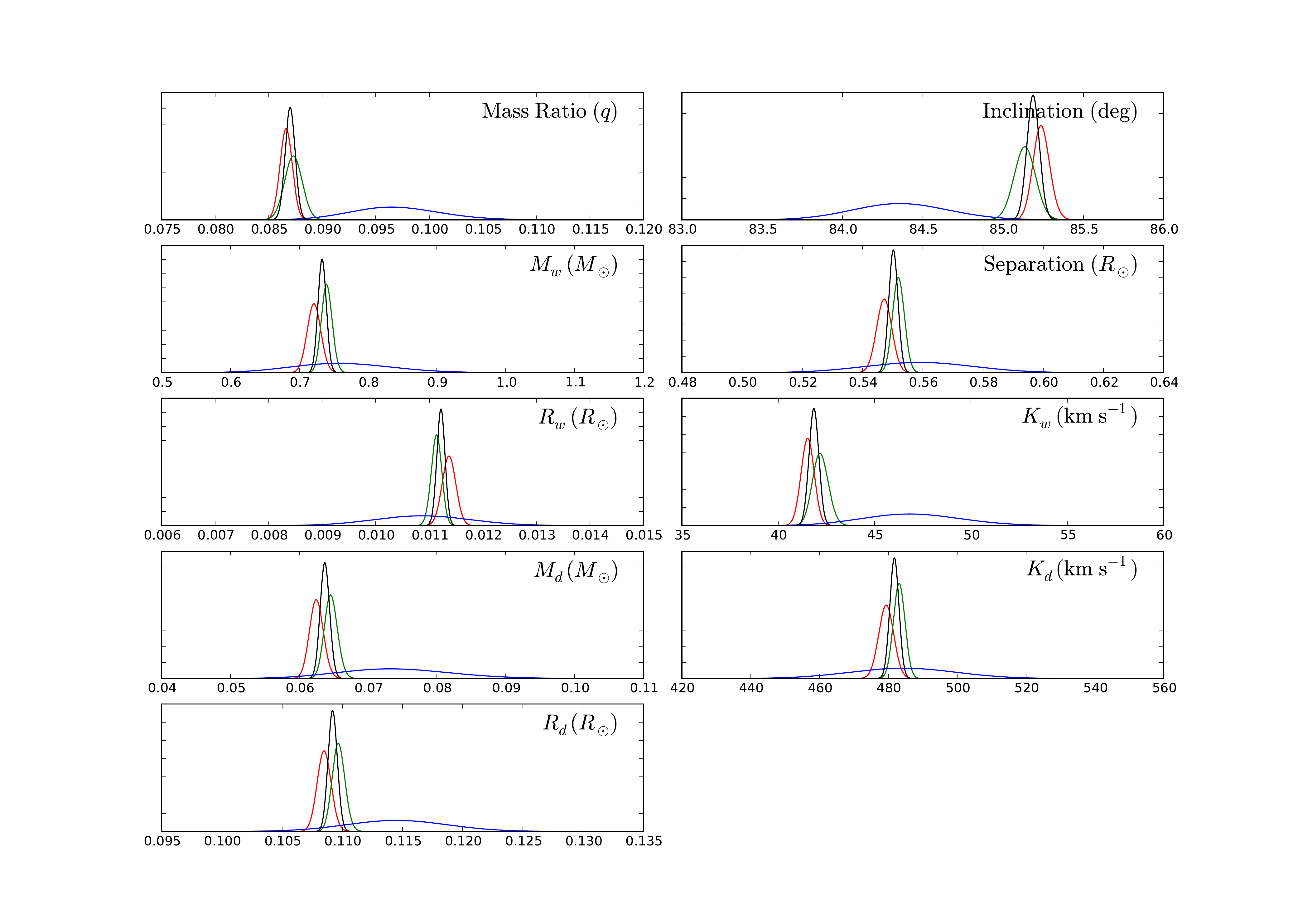}
\caption{\label{fig:pdf} Normalised posterior probability density functions (black) for each parameter of the model. The red, green and blue distributions represent the $r'$, $g'$ and $u'$ band fits, respectively.}
\end{center}
\end{figure}

\begin{figure*}
\begin{center}
\includegraphics[width=1.0\columnwidth,trim=300 30 300 30]{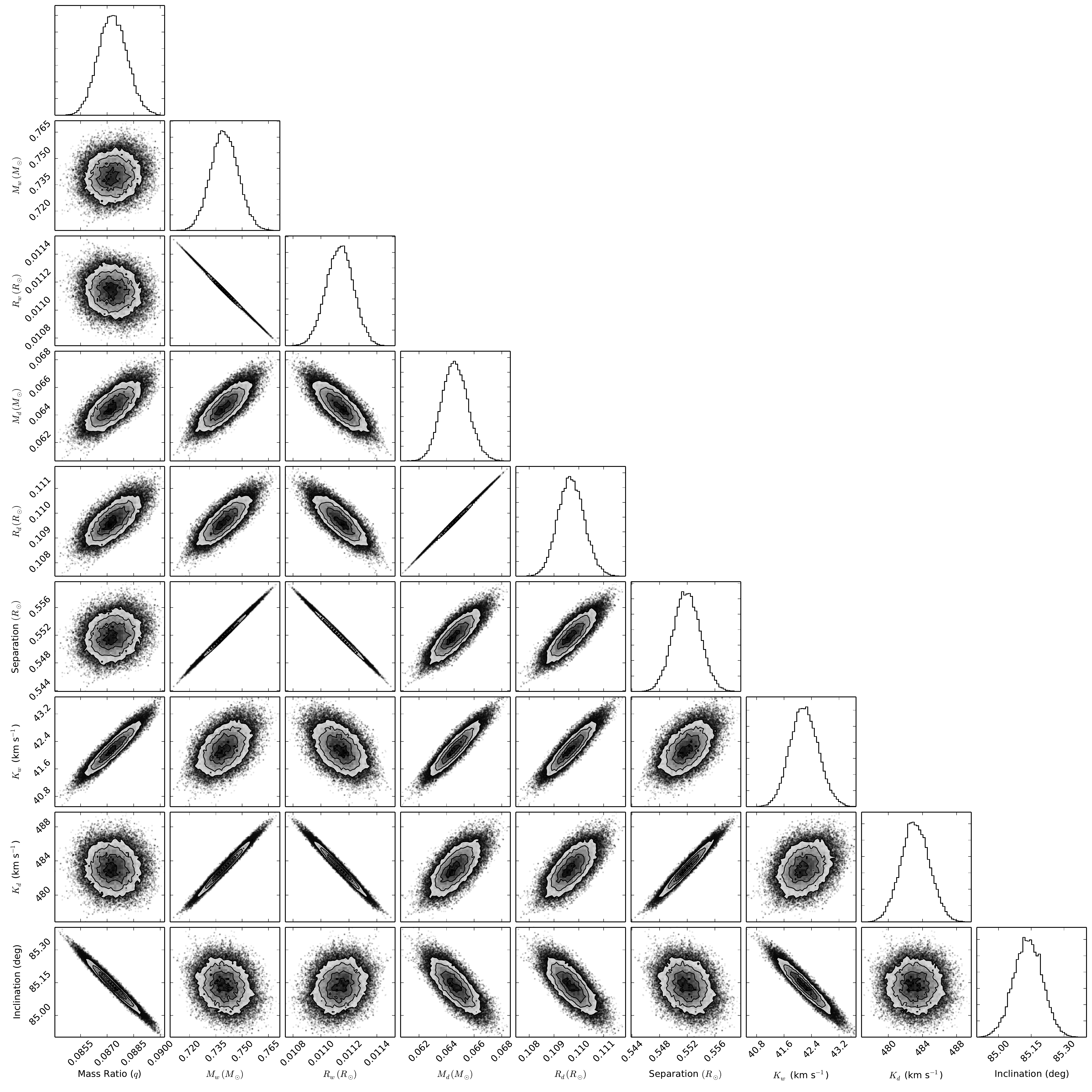}
\caption{\label{fig:cornerplot} Corner plot of $g'$ band fit showing correlations of varying strengths between system parameters.}
\end{center}
\end{figure*}

Combining the posterior probability distributions from the $u'$, $g'$ and $r'$ bands gave the total posterior distributions for each system parameter (Figure~\ref{fig:pdf}), with the peak of this distribution taken as the value of that particular system parameter. Upper and lower error bounds are derived from the 67\% confidence levels. Figure~\ref{fig:cornerplot} shows a corner plot for the $g'$ band fit, which exposes degeneracies between certain system parameters.

The system parameters were calculated twice in total. The value for log\,$g$ returned from the first calculation was used to constrain the log\,$g$ prior in a second MCMC fit involving the white dwarf atmosphere predictions \citep{bergeron95}. This second MCMC fit also used white dwarf fluxes from four wavelength bands instead of the three used previously. The additional $i'$-band white dwarf flux was obtained through fitting the eclipse model to the individual $i'$-band eclipse from the Sep 2012 observation (see section~\ref{subsec:indlcmod} for more details on this and other individual eclipse fitting). A systematic error of 3\% had to be added to the fluxes in order to reach a $\chi^{2}$ of $\sim$1, which is of the same order as the out-of-eclipse photometric error ($\sim$2\%) and approximately half that of the error associated with flickering ($\sim$5\%). The use of an additional bandpass and a constraint on log\,$g$ resulted in more precise values for the white dwarf temperature and distance. This new temperature was then used to obtain a more reliable white dwarf mass-radius relationship, which was used in the second calculation of the system parameters.

\begin{figure}
\begin{center}
\includegraphics[width=1.0\columnwidth,trim=40 20 50 20]{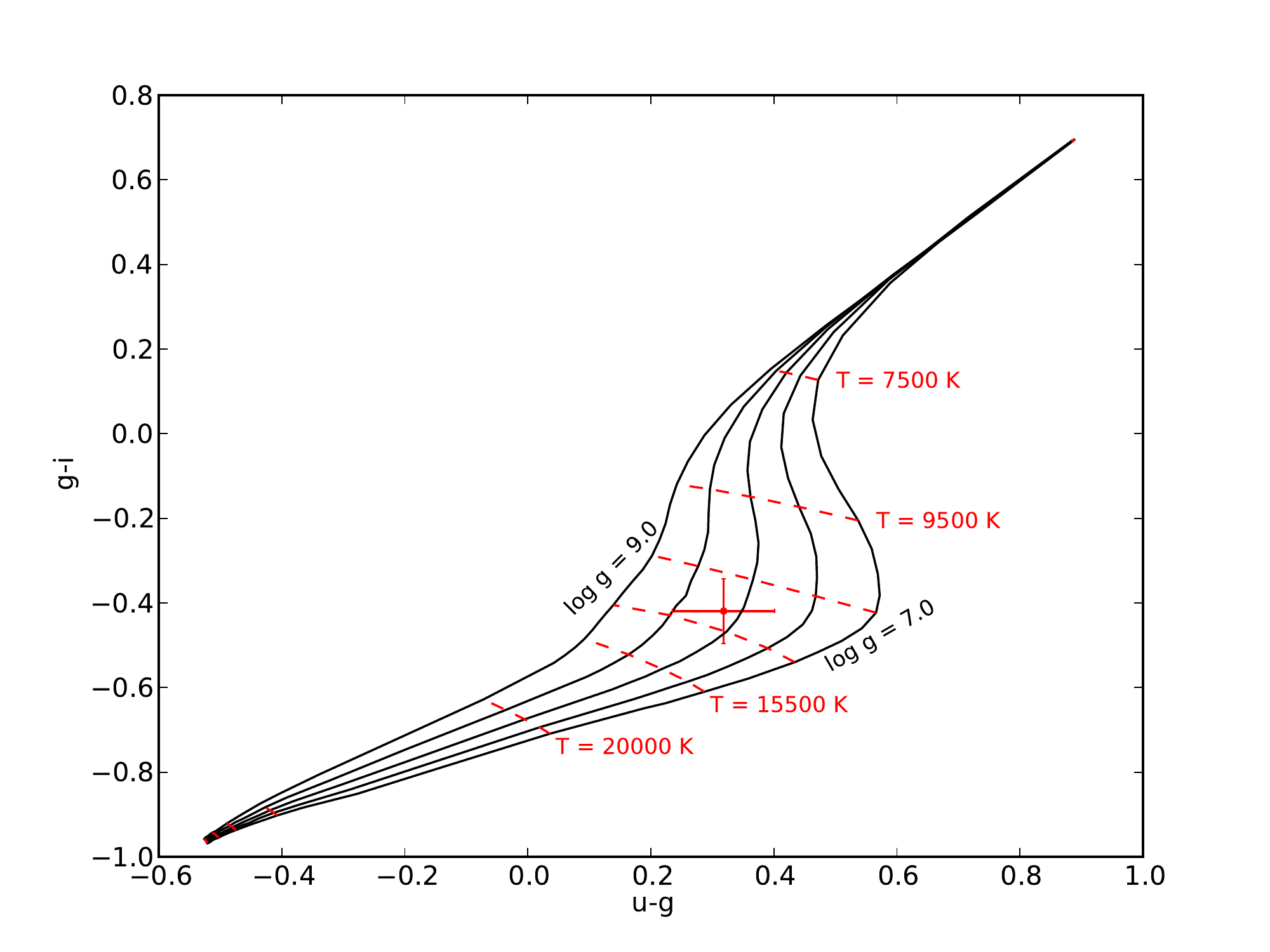}
\caption{\label{fig:colourplot}White dwarf colour-colour plot. The PHL 1445 white dwarf is shown by the red data point and the black lines represent white dwarf models of varying log\,$g$ from \protect\cite{bergeron95}.}
\end{center}
\end{figure}

Figure~\ref{fig:colourplot} shows a white dwarf colour-colour plot, containing both the colour of the white dwarf in PHL 1445 and models from \cite{bergeron95}. As expected, there is good agreement between the colour of the white dwarf and the temperature and log\,$g$ values determined from fitting to these models.

The calculated system parameters can be found in Table~\ref{table:params}. The errors in the first three columns of Table~\ref{table:params} are those resulting from the MCMC fitting only, and do not account for uncertainties related to the assumptions associated with the model (see \citealt{savoury11}) or those arising from the effects of flickering. Flickering affects the system parameters because it decreases the accuracy to which the eclipse timings - especially bright spot ingress \& egress - can be measured. As this particular system displays strong flickering, it is clear that the errors on the system parameters from the model are underestimated, even though multiple lightcurves have been averaged.

To probe the effects of flickering, four additional $g'$ band average lightcurves were produced and fit with the eclipse model. Each of these new lightcurves contained a different combination of three of the four individual lightcurves used in the original $g'$ band average lightcurve. The spread of system parameters obtained from these four MCMC fits gives a more realistic idea of the errors involved; the errors in the final column of Table~\ref{table:params} include our estimate of the uncertainty introduced by flickering.

\begin{table*}
\begin{center}
\begin{tabular}{lcccccc}
\hline
Band & $r'$ & $g'$ & $u'$ & Combined & Sys err (\%) due & Final \\ 
&&&&& to flickering & estimates\\ \hline
$q$ & $0.0866 \pm 0.0006$ & $0.0873 \pm 0.0008$ & $0.096 \pm 0.004$ & $0.08701 \pm 0.0005$ & 7 & $0.087 \pm 0.006$ \\
$M_{w}$ ($M_{\odot}$) & $0.722 \pm 0.010$ & $0.740 \pm 0.008$ & $0.76 \pm 0.08$ & $0.733 \pm 0.006$ & 4 & $0.73 \pm 0.03$ \\
$R_{w}$ ($R_{\odot}$) & $0.01137 \pm 0.00013$ & $0.01114 \pm 0.00010$ & $0.0109 \pm 0.0009$ & $0.01122 \pm 0.00008$ & 3 & $0.0112 \pm 0.0003$ \\
$M_{d}$ ($M_{\odot}$) & $0.0625 \pm 0.0010$ & $0.0646 \pm 0.0010$ & $0.073 \pm 0.009$ & $0.0637 \pm 0.0007$ & 7 & $0.064 \pm 0.005$ \\
$R_{d}$ ($R_{\odot}$) & $0.1085 \pm 0.0006$ & $0.1097 \pm 0.0006$ & $0.115 \pm 0.004$ & $0.1092 \pm 0.0004$ & 4 & $0.109 \pm 0.004 $ \\
$a$ ($R_{\odot}$) & $0.547 \pm 0.003$ & $0.552 \pm 0.002$ & $0.559 \pm 0.018$ & $0.5502 \pm 0.0016$ & 2 & $0.550 \pm 0.011$ \\
$K_{w}$ (km\,s$^{-1}$) & $41.5 \pm 0.3$ & $42.2 \pm 0.4$ & $47 \pm 3$ & $41.8 \pm 0.3$ & 6 & $42 \pm 3$ \\
$K_{r}$ (km\,s$^{-1}$) & $479 \pm 2 $ & $483.1 \pm 1.7$ & $484 \pm 16$ & $481.7 \pm 1.4$ & 1 & $482 \pm 5$ \\
$i$ $(^{\circ})$ & $85.24 \pm 0.05$ & $85.14 \pm 0.07$ & $84.4 \pm 0.3$ & $85.19 \pm 0.04$ & 1 & $85.2 \pm 0.9$ \\ \hline
log\,$g$ &&&& $8.2 \pm 0.3$ && $8.2 \pm 0.3$ \\
$T_{w}$ (K) &&&& $13200 \pm 700$ && $13200 \pm 700$ \\
$d$ (pc) &&&& $220 \pm 50$ && $220 \pm 50$ \\
\hline
\end{tabular}
\caption{\label{table:params}System parameters for PHL 1445. The errors in the combined column are returned by the model and are purely statistical. The errors in the final column take into account the systematic error due to flickering. $T_{w}$ and $d$ represent the temperature and distance of the white dwarf, respectively.}
\end{center}
\end{table*}

\subsection{Individual lightcurve modelling}
\label{subsec:indlcmod}

\begin{figure*}
\begin{center}
\includegraphics[width=1.7\columnwidth,trim=0 30 0 30]{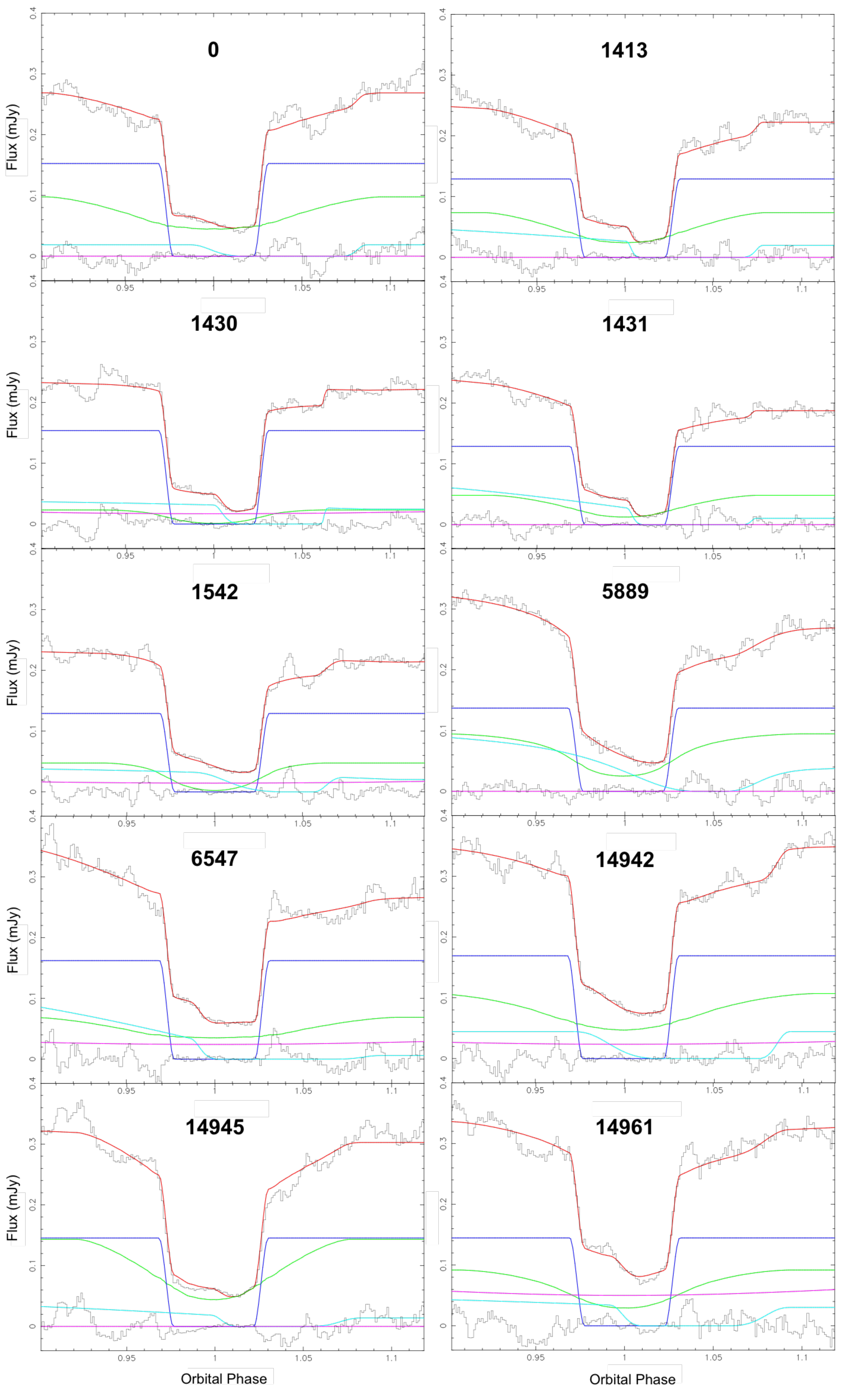}
\caption{\label{fig:indfits}Model fits (red) to individual PHL 1445 $g'$-band eclipses (black). The additional coloured lines are explained in Figure~\ref{fig:avefits}. The cycle numbers of each eclipse are also displayed.}
\end{center}
\end{figure*}

After determining various system parameters using an average eclipse lightcurve, the eclipse model could now be fit to individual lightcurves (as long as they showed signs of a bright spot ingress), using the model fit parameters as a starting point. The eclipse model parameters $q$, $\Delta\phi$, $R_{w}/a$ and $U_{w}$ don't vary with time, so these parameter values were kept fixed in the individual fits.

In total there were 10 eclipses that showed signs of a bright spot ingress feature, and therefore qualified for individual MCMC fitting, including the four used in the phase-folded average fitting. All but one of these eclipses was observed in the wavelength bands $u'g'r'$; the other in $u'g'i'$. Each individual eclipse was fit in each of the three bands, with the starting model parameters depending on the band. The one $i'$-band eclipse (08 Sep 2012) was given the overall $g'$-band model parameters as a starting point in the MCMC fitting, but using an $i'$-band $U_{w}$ value of 0.301 \citep{gianninas13}. Successful fits to bright spot ingress were achieved for all 10 eclipses. Since $q$ and $\Delta\phi$ were held fixed, and bright spot ingress/egress timings are functions of $q$, $\Delta\phi$ and radius of the accretion disc as a fraction of the binary separation ($R_{\rm disc}/a$), $R_{\rm disc}/a$ could be constrained for all 10 eclipses.

The individual $g'$-band lightcurves and corresponding eclipse model fits are shown in Figure~\ref{fig:indfits}. Individual eclipse fitting should enable us to analyse how various parameters vary from eclipse to eclipse, for example disc radii and component fluxes. However, due to the strong flickering present in each lightcurve, it is important to check these fits are genuine and interpret the results with care. Looking at the individual fits in Figure~\ref{fig:indfits}, it is clear that not all achieve a true fit to the bright spot features (e.g. cycle numbers 5889, 14942 \& 14945), and this will be taken into account in the following discussion.

The individual eclipse fitting carried out on PHL 1445 provided nine separate sets of $u'g'r'$ fluxes for the white dwarf, accretion disc and bright spot. Following section~\ref{subsec:syspars}, a systematic error of 3\% was added to all fluxes returned by the individual fits. There was no evidence for a varying white dwarf temperature across these nine observations.

\subsubsection{Accretion disc}
\label{subsubsec:accdisc}

\begin{figure}
\begin{center}
\includegraphics[width=1.0\columnwidth,trim=110 10 110 60]{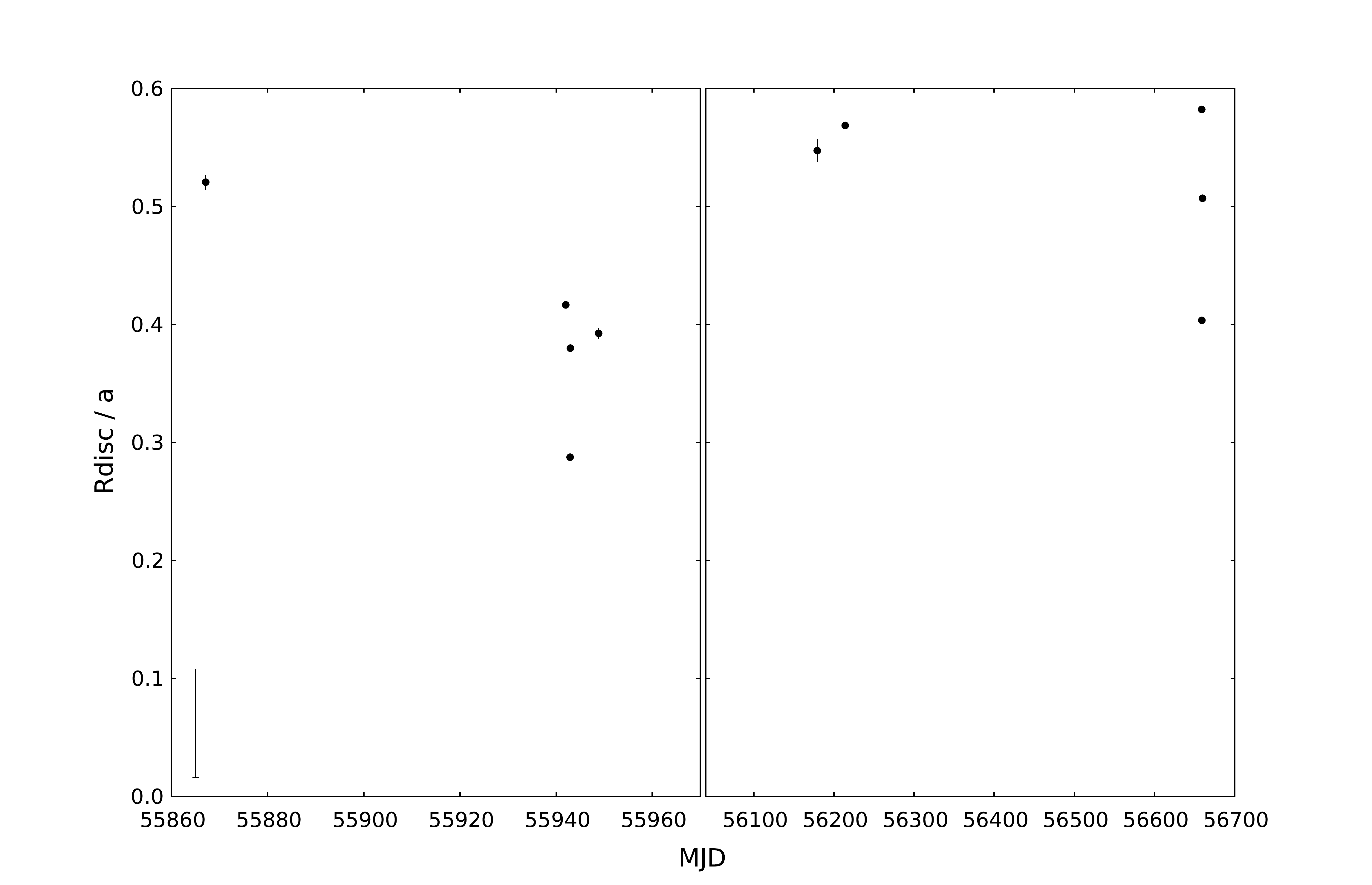}
\caption{\label{fig:discradvstime}PHL 1445 accretion disc radius ($R_{disc}$) as a fraction of the binary separation ($a$) vs time (in MJD). Individual errors are purely statistical, and in most cases the error bars are smaller than the data points. The bar in the bottom-left corner gives an indication of the real error due to flickering. The figure is split into two due to a large gap in time ($\sim$200\,days) between observations.}
\end{center}
\end{figure}

Individual eclipse fitting produced a value of $R_{disc}/a$ for all 10 eclipses. This value from the model is actually the bright spot's distance from the white dwarf as a fraction of the binary separation, but we assume the bright spot is lying at the edge of the accretion disc. With the $u'$-band fits being the least reliable due to the low quality of lightcurves, only the the $r',i'$- and $g'$-band $R_{disc}/a$ were used. For each eclipse, an average of the $r',i'$- and $g'$-band $R_{disc}/a$ was plotted against $T_{mid}$ to show how the disc varies with time, as shown in Figure~\ref{fig:discradvstime}. The plot is split into two due to a sizeable time gap between observations. 

The individual errors displayed in Figure~\ref{fig:discradvstime} are from the model fits, and are dramatically underestimated due to the effects of flickering. There is a systematic error on the disc radius of approximately 10\% due to flickering, and this is represented by the bar in the bottom-left corner of Figure ~\ref{fig:discradvstime}. Without the introduction of this systematic error the disc changes appear to be very large, for example take the successive eclipses of 1430 and 1431. In the time of just one orbital period (76.3\,mins), $R_{disc}/a$ increases from approximately 0.288 to 0.380, implying a disc expansion velocity of $\rm7200\,m\,s^{-1}$, which is significantly faster than the $\rm2\,m\,s^{-1}$ ``viscous velocity'' of material within the disc. As Figure~\ref{fig:indfits} shows, both of these eclipses have clear bright spot ingress features. The fit to the bright spot ingress in the 1431 eclipse is far better than that in the 1430 eclipse, and this may be the real reason for the large disc radius expansion observed over this orbital cycle. The poor fit to the bright spot ingress in cycle 1430, and in many other individual eclipses (Figure~\ref{fig:indfits}), is most likely due to the large amount of flickering, which we address with the introduction of a 10\% systematic error. It must be noted that in some individual eclipses with weak bright spot features (e.g. cycle numbers 1542 \& 5889) the bright spot ingress is hardly fit at all, resulting in much more uncertain values of $R_{disc}/a$ in these cases. 

%The most significant change in disc radius occurs between the successive eclipses observed on 16 Jan 2012 (cycle numbers 1430 \& 1431). In the time of just one orbital period (76.3\,mins), $R_{disc}/a$ increases from approximately 0.288\,$R_{\odot}$ to 0.380\,$R_{\odot}$, implying a disc expansion velocity of $\rm7200\,m\,s^{-1}$. Two velocities to compare this to are the sound speed through the disc and the ``viscous velocity'' of material within the disc. The expansion velocity is found to be lower than that of sound through the disc as long as disc temperature is greater than 4000\,K, which is most likely the case. The viscous velocity through the disc at a temperature as high as 10000\,K is $\rm2\,m\,s^{-1}$, much lower than the expansion velocity. Of course, this apparently large change in disc radius may be dominated by systematic errors due to flickering.

\begin{figure*}
\begin{center}
\includegraphics[width=1.0\columnwidth,trim=150 30 55 40]{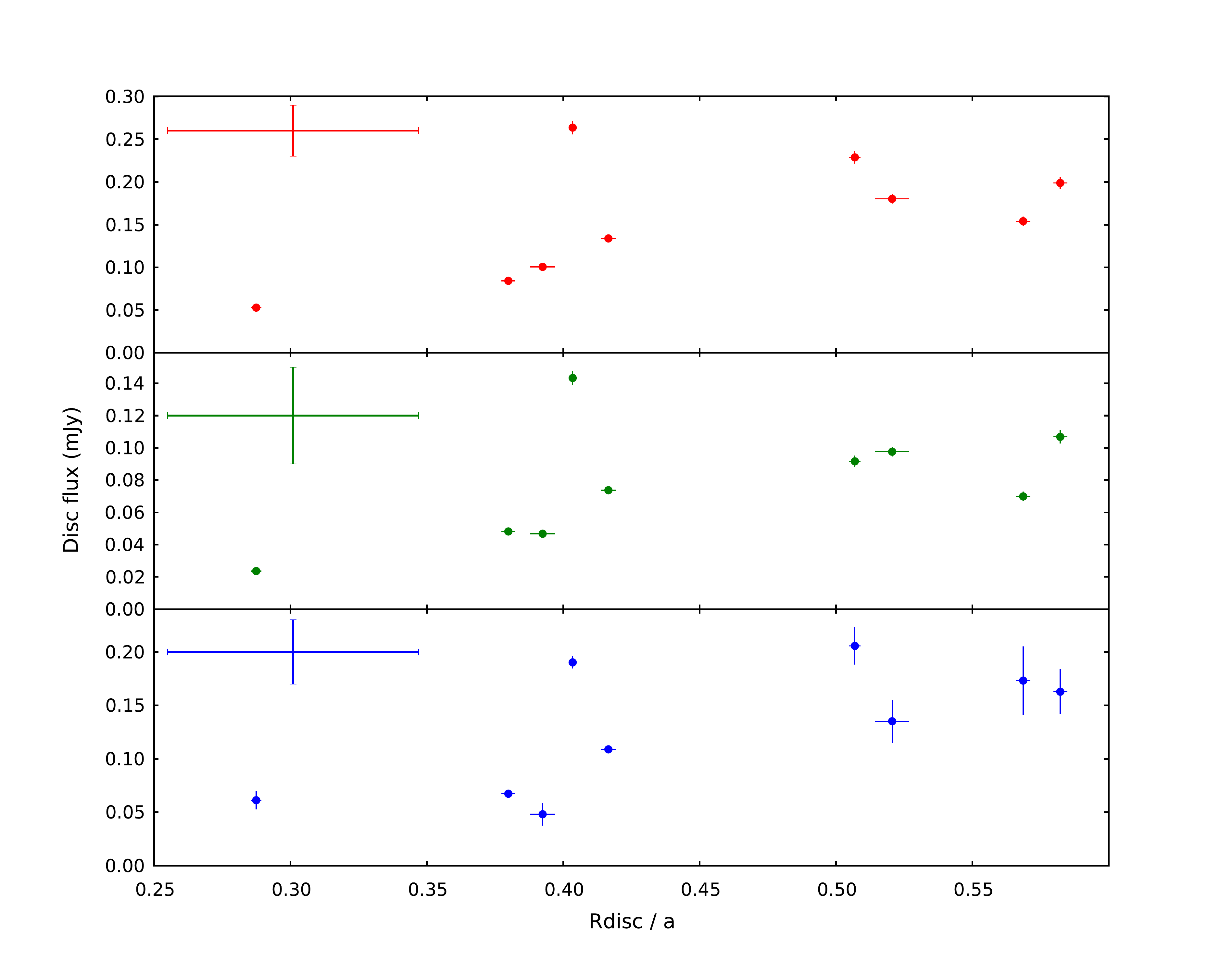}
\includegraphics[width=1.0\columnwidth,trim=65 30 150 40]{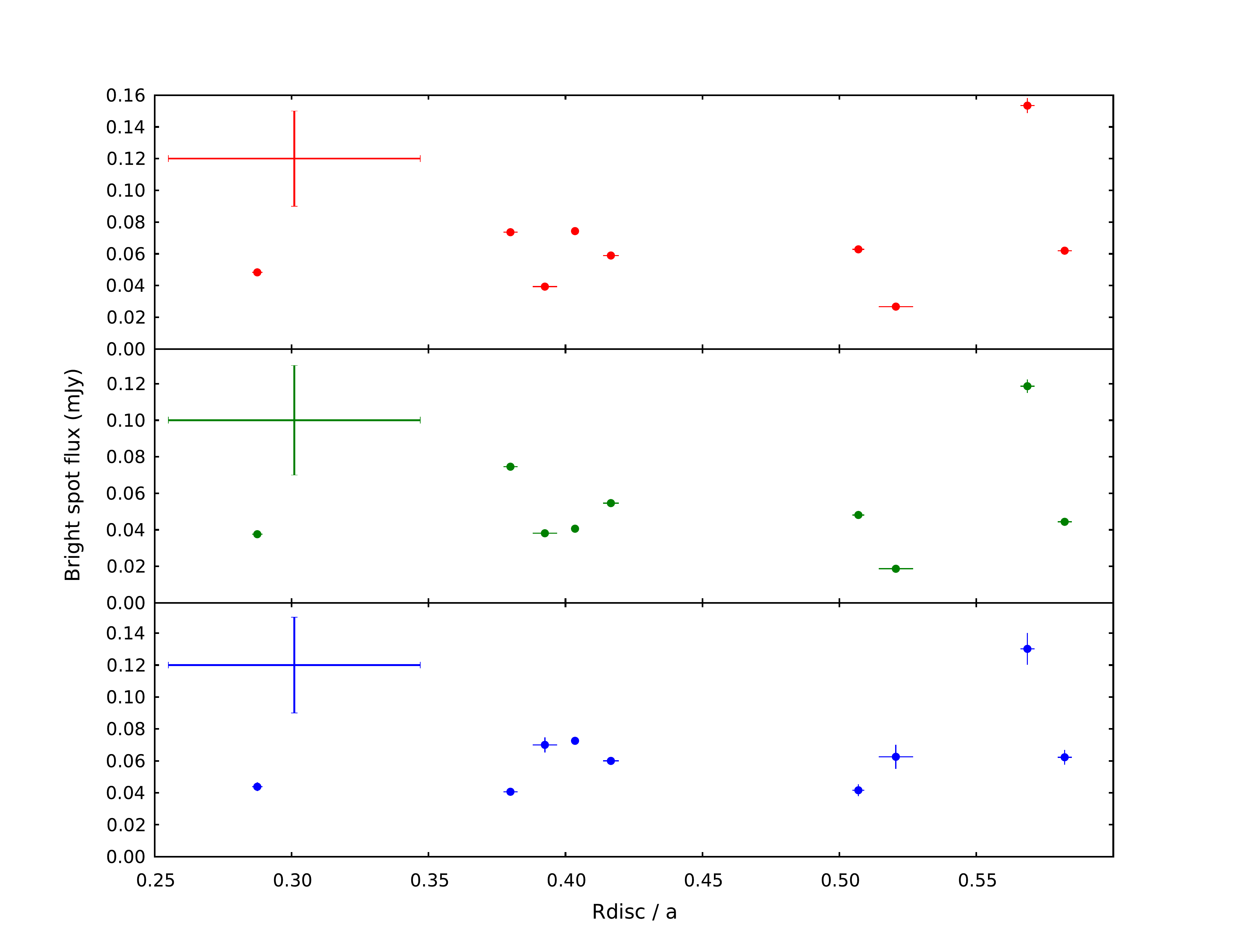}
\caption{\label{fig:discradvsdiscflux/bsflux}PHL 1445 accretion disc flux (left) and bright spot flux (right) vs radius ($R_{disc}$) as a fraction of the binary separation ($a$). Red, green and blue data points represent $r'$, $g'$ and $u'$ band observations, respectively. Individual errors are purely statistical, and in most cases the error bars are smaller than the data points. The bars in the top-left corners of each plot give an indication of the real error due to flickering.}
\end{center}
\end{figure*}

The left plot in Figure~\ref{fig:discradvsdiscflux/bsflux} shows how the flux of the disc varies with disc radius. Again, the individual errors are underestimated, and the errors introduced by flickering are represented by the bars in the top-left corner of each plot. To measure the reliability of these flux changes, we again turn to the successive eclipse cycles 1430 \& 1431. Here we find that in one orbital period the disc flux increases by $\sim\rm0.025\,mJy$, but at the same time the white dwarf flux drops by the same amount. Such a change in white dwarf flux over one orbital cycle is not expected, and it is clear by looking at Figure~\ref{fig:indfits} that a fraction of the white dwarf flux in cycle 1430 has in the following cycle been fit by the disc component instead. This may not be the case for all individual eclipse fits, but it does question the reliability of the model disc flux values. The most likely cause for this is the large amount of flickering in these individual eclipse lightcurves, which confirms the need for a large systematic error to account for it. Despite the large errors, there does appear to be a positive correlation between these two disc parameters. There is no evidence for changes in disc temperature, so the trend in Figure~\ref{fig:discradvsdiscflux/bsflux}, if real, appears to be simply due to a larger disc radius resulting in a larger disc surface area and therefore flux. 

\subsubsection{Bright spot}
\label{subsubsec:bs}

The bright spot fluxes were also plotted against $R_{disc}/a$ (see right plot in Figure~\ref{fig:discradvsdiscflux/bsflux}). Unlike the disc fluxes, on the whole the bright spot fluxes appear to stay relatively constant across the different disc radii. Assuming the main contributor to bright spot flux is the relative velocity of the gas stream as it impacts with the disc, we modelled the gas stream and calculated its velocity relative to the disc for a number of disc radii across the range 0.2 \textless\,$R_{disc}/a$\,\textless\,0.6 . The relative velocity of the gas stream only increased by a factor of two across this range, which could explain why we see little variation in bright spot fluxes.

An attempt was also made to determine bright spot temperatures from the bright spot fluxes. The bright spot fluxes in each band from each eclipse were used to calculate bright spot colours, and then compared to local thermal equilibrium (LTE) hydrogen slab models calculated using {\sc synphot} in {\sc iraf}. Due to the large error bars on the eclipse fluxes, and the rapid changes in colour with variations in temperature \& density associated with the models, accurate bright spot temperatures could not be determined in this particular case.

\section{Discussion}
\label{sec:discussion}

\subsection{Component masses}
\label{subsec:wdmass}

We find a white dwarf mass of $0.73 \pm 0.03$\,$M_{\odot}$ in PHL 1445, which is much larger than that of white dwarfs in single and pre-CV systems \citep{zorotovic11,kepler07}, but is identical to the average white dwarf mass in CV systems below the period gap ($0.73 \pm 0.05$\,$M_{\odot}$) found by \cite{knigge06}. It is however lower than the mean white dwarf mass found by \cite{savoury11} within a group of 14 short period CVs ($0.81 \pm 0.04$\,$M_{\odot}$). We expect this mass for the white dwarf in PHL 1445 to be reliable, as previous mass determinations using this method agree with those obtained through spectroscopic methods \citep{tulloch09,savoury12,copperwheat12}.

As for the donor in PHL 1445, we find it has a mass of $0.064 \pm 0.005\,M_{\odot}$. This is just below the hydrogen burning mass limit of $\sim$0.075\,$M_{\odot}$ \citep{kumar63,hayashi63}, which suggests it is substellar.

\subsection{Flickering}
\label{subsec:flickering}

One particular feature of the PHL 1445 eclipses is high amplitude flickering. This is much larger than the flickering observed in other known CVs with substellar donors (see section~\ref{subsec:periodbouncer}), and appears to be originating from the inner accretion disc, near the white dwarf. The accretion disc and bright spot fluxes as a fraction of the white dwarf flux were calculated in $u'$, $g'$ and $r'$, and compared with those from other CVs with substellar donors. PHL 1445's disc fluxes are nearly double those of the second highest disc flux system, which suggests the enhanced flickering is associated with a brighter disc.

\subsection{Evolutionary state of PHL 1445}
\label{subsec:evostate}

It is known that PHL 1445 consists of an accreting white dwarf and a donor star, but what is not known is the nature of this donor star. It is possible that PHL 1445 lies below the period minimum because it contains an unusual donor star, one most likely off the main sequence. We have determined a mass, radius and flux for the donor star of PHL 1445, allowing us to investigate this.

One possibility is that PHL 1445 is a Galactic halo object. A system belonging to the Galactic halo would typically have a metal-poor donor star, meaning a smaller than expected radius for its mass and therefore a higher density. Due to the inverse relation between density of a Roche lobe-filling donor star and the orbital period of a system, a metal-poor donor is one way for a CV system to have an orbital period below the period minimum \citep{patterson08}. This was found to be the case for SDSS J150722.30+52309.8 (SDSS 1507), another CV with an orbital period (67\,mins) below the minimum \citep{patterson08,uthas11}.

SDSS 1507's halo membership is supported by both its unusually high space velocity ($\rm167\,km\,s^{-1}$), calculated from its distance \& proper motions by \cite{patterson08} and sub-solar metallicity determined from UV spectroscopy by \cite{uthas11}. Using our distance to PHL 1445 and proper motions listed in the PPMXL catalog \citep{roeser10}, a transverse velocity of $\rm 39 \pm 9\,km\,s^{-1}$ was calculated. This is significantly lower than that for SDSS 1507, and is very close to the average transverse velocity of $\rm 33\,km\,s^{-1}$ for CVs \citep{patterson08}, which is evidence against PHL 1445 being a member of the Galactic halo.
 
Another explanation for the short orbital period is a donor star that is already evolved at the start of mass transfer. One way of determining the evolutionary stage of a star is through its composition, which can be determined from its spectrum. A spectrum for PHL 1445 is shown in \cite{wils09}, but this isn't useful to us as it's dominated by the other components of the CV, not the donor. This isn't surprising, as we also fail to directly detect the donor (see Figure~\ref{fig:avefits}). Through model fitting we do obtain an upper limit for the donor flux, which is actually a measure of the total un-eclipsed flux from the system.

\cite{thorstensen02} show that an evolved donor with a central hydrogen abundance of $X_{c}$ = 0.05, in a system with PHL 1445's orbital period should have a temperature in excess of 4000\,K, while \cite{podsiadlowski03} show an $X_{c}$ = 0.1 evolved donor in a similar system to have a temperature somewhere between 1500-2000K. Through knowledge of PHL 1445's donor angular diameter and flux from the eclipse model, we are able to rule out a 4000\,K donor, but do find some agreement with a 1500-2000\,K donor. 

$g'$\,-\,$r'$ colours were estimated for both a 4000\,K and 1800\,K donor. The colour for the 4000\,K donor was found through the linear relation between $T_{\rm eff}$ and $g'$\,-\,$r'$ \citep{fukugita10}, but this relation doesn't extend to below $\sim$3800\,K so semi-empirical model isochrones had to be used in order to obtain a colour for the 1800\,K donor \citep{baraffe98,allard11,bell14}. From these colours, $r'$-band zero-magnitude angular diameters were calculated and used together with the donor angular diameter to produce an apparent $r'$-band donor magnitude at each temperature \citep{boyajian14}. It must be noted that the colour obtained for the 1800\,K donor lies outside the valid range given by \cite{boyajian14} for their magnitude-angular diameter relation.

Donor fluxes of ($15.9 \pm 1.1) \times 10^{-2}$\,mJy \& ($0.33 \pm 0.18) \times 10^{-2}$\,mJy were calculated for 4000\,K \& 1800\,K, respectively. The 4000\,K donor is approximately 13 times the ($1.27 \pm 0.08) \times 10^{-2}$\,mJy upper limit for the $r'$-band donor flux from the eclipse model, while the 1800\,K donor flux is approximately 4 times smaller. Analysis of the donor flux hence shows that a slightly evolved donor ($X_{c}$ = 0.1) cannot be ruled out for PHL 1445, and may be the reason for its unusually short orbital period.

\begin{figure*}
\begin{center}
\includegraphics[width=1.5\columnwidth,trim=80 20 80 30]{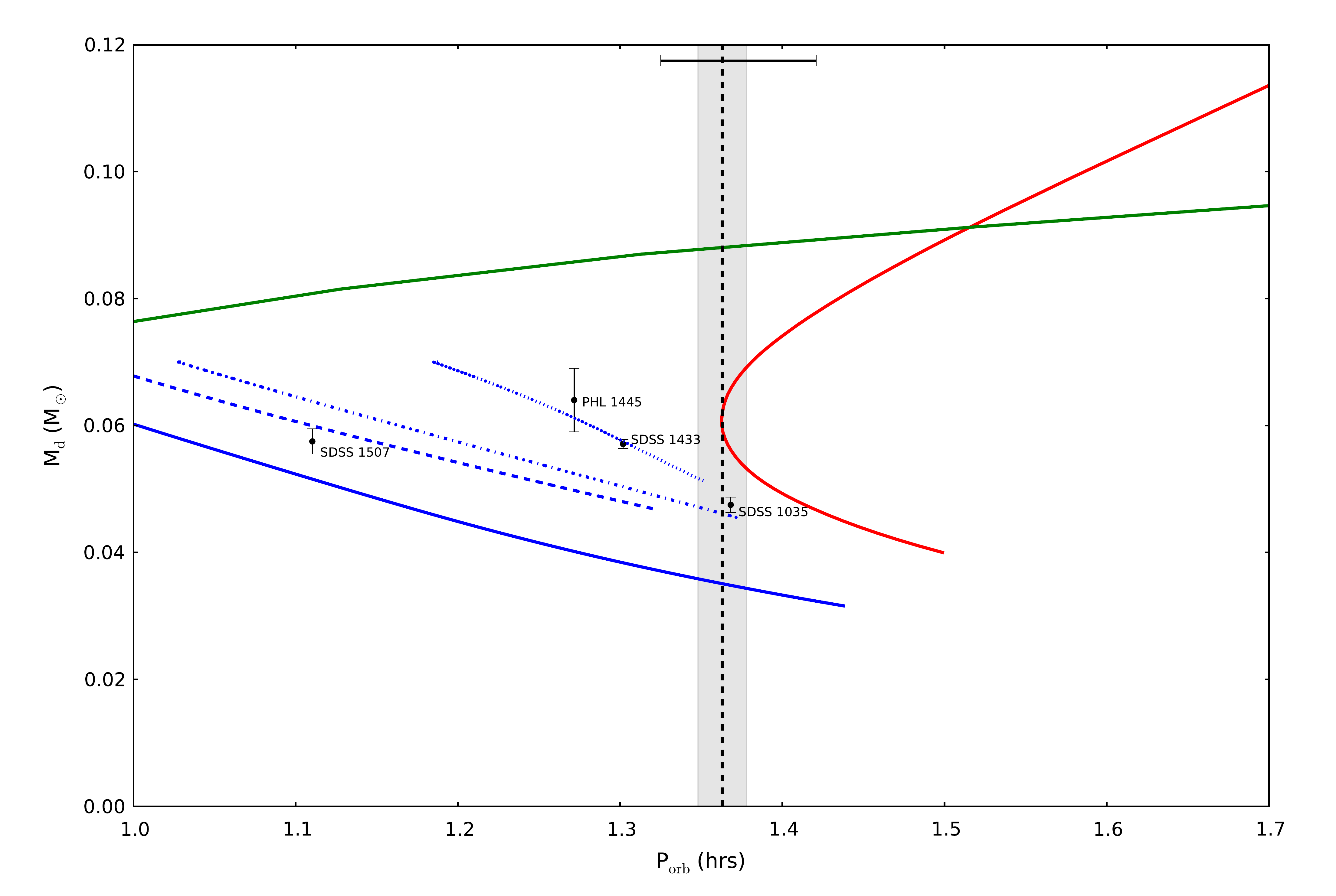}
\caption{\label{fig:m2vsp}Donor mass ($M_{d}$) vs orbital period ($P_{\rm orb}$) for PHL 1445 and other substellar donor CVs: SDSS 1433, SDSS 1035 \& SDSS 1507 \citep{savoury11}. Also plotted are a number of evolutionary tracks: main sequence donor (red line; \protect\citealt{knigge11}), evolved donor with $X_{c}$ = 0.05 (green line; \protect\citealt{thorstensen02}) and brown dwarf donor (solid blue line; \protect\citealt{kolbbaraffe99}). The three additional blue lines also show tracks for brown dwarf donors but with modified physics and varying donor age at start of mass transfer ($t_{init}$). The dashed, dot-dashed and dotted blue lines represent $t_{init}$'s of 2\,Gyr, 1\,Gyr and 600\,Myr, respectively. The vertical dashed black line represents the location of the CV period minimum determined by \protect\cite{knigge11}, with the shaded area representing the error on this value. The bar across the top of the plot shows the FWHM of the CV period spike observed by \protect\cite{gaensicke09}.}
\end{center}
\end{figure*}

PHL 1445 could also lie below the period minimum because it formed directly with a brown dwarf donor. These systems can start out with periods much shorter than the period minimum, but evolve towards longer orbital periods like post-period bounce CVs \citep{kolbbaraffe99}. We investigate whether PHL 1445 could have formed with a brown dwarf donor by studying the relation between donor mass and orbital period (see Figure~\ref{fig:m2vsp}). Figure~\ref{fig:m2vsp} shows a number of different evolutionary tracks. The red track is from \cite{knigge11} and represents a CV with a main sequence donor. These CVs evolve from longer periods to shorter ones until the period minimum (vertical dashed line) is reached, at which point the track inverts and heads back to longer periods. The green track is from \cite{thorstensen02} and represents a system containing an evolved donor with $X_{c}$ = 0.05. Above we rule out the possibility of such a highly-evolved donor, and this is supported by the fact that the PHL 1445 data point lies comfortably below this line. The solid blue line is from \cite{kolbbaraffe99} and represents a system that formed with a brown dwarf donor. It would appear that the PHL 1445 data point lies far from this track, but this track is computed from an old model, using a gravitational-radiation (GR) based angular momentum loss rate and ignoring deformation of the donor. \cite{knigge11} showed that tracks with these assumptions cannot fit the observed locus of CVs in the $M_{d}$ vs $P_{\rm orb}$ diagram, and that models which include deformation and an angular momentum loss rate of 2.47xGR are required. The main sequence donor track (red) in Figure~\ref{fig:m2vsp} takes into account both a 2.47xGR angular momentum loss rate and deformation \citep{renvoize02}, and for the additional brown dwarf donor tracks (blue: dashed, dot-dashed and dotted) we have done the same. All three of these tracks have been calculated from a model containing a 0.75\,$M_{\odot}$ primary and a donor of initial mass 0.07\,$M_{\odot}$, with an additional variable parameter being the age of the donor at start of mass transfer ($t_{init}$). This is an important parameter with regards to understanding the subsequent evolution of such a system, since a substellar object has a time-dependent radius. The dashed, dot-dashed and dotted blue lines represent  $t_{init}$'s of 2\,Gyr, 1\,Gyr and 600\,Myr respectively. The latter of these tracks - with a $t_{init}$ of 600\,Myr - is consistent with the PHL 1445 data point, but how feasible is such a proposed system?

In order for mass transfer to start so early in the system's lifetime, the primary star must have evolved off the main sequence very quickly to leave a white dwarf ready for mass transfer. This puts a lower limit on the initial primary mass of 2.8\,$M_{\odot}$ \citep{girardi00}. Considering the secondary has an initial mass no greater than 0.07\,$M_{\odot}$, this would mean an initial mass ratio of approximately 0.025 or less. This is extremely low, and main sequence star/brown dwarf binaries with extreme mass ratios are rare, but it would seem they are able to form \citep{grether06}. Binaries with such low mass ratios have been observed, for example HIP 77900B, which has a mass ratio as low as 0.005 \citep{aller13}, although its separation is also extreme at 3200\,AU.  There is also observational evidence for binaries with A-type star primaries to have a bias towards low mass ratios of less than 0.1 \citep{Kouwenhoven05}. It is thus a possibility that PHL 1445 formed directly from a binary system with a very low mass ratio containing a $\textgreater$2.8\,$M_{\odot}$ primary and brown dwarf secondary.

\subsection{PHL 1445: a possible period-bouncer?}
\label{subsec:periodbouncer}

Another CV, SDSS J143317.78+101123.3 (SDSS 1433) has also been found to host a substellar donor \citep{littlefair08,savoury11}. SDSS 1433 - with an orbital period of 78.1\,mins and donor mass of $0.0571 \pm 0.0007$\,$M_{\odot}$ - was claimed to be a post-period bounce system by both \cite{littlefair08} and \cite{savoury11}. Current estimates now place the period minimum at $81.8 \pm 0.9$\,min \citep{knigge11}, suggesting that SDSS 1433 may not, in fact, be a period bouncer.

With an orbital period below the period minimum and a substellar donor, SDSS 1433 may be another system that formed with a brown dwarf donor. This is supported by plotting SDSS 1443 on the same $M_{d}$ vs $P_{\rm orb}$ diagram as PHL 1445 (see Figure~\ref{fig:m2vsp}). SDSS 1433 also appears to lie on the evolutionary track associated with a brown dwarf donor of age 600\,Myr at start of mass transfer, and therefore may have had a progenitor system with the same, extremely low mass ratio as that of PHL 1445. The chances of finding two systems with such similar, extreme initial mass ratios should be very low, so the fact we do raises some suspicion. 

There are currently only four CV systems that have been found to contain substellar donors: PHL 1445, SDSS 1433, SDSS 1507 and SDSS J103533.03+055158.4 (SDSS 1035) \citep{littlefair08,savoury11}. SDSS 1507 was mentioned in the previous section, and lies below the period minimum due to being a member of the Galactic halo. SDSS 1035 has an orbital period slightly above the period minimum (82.1\,min, \citealt{savoury11}), and is thought to be a period-bouncer. So out of a sample of just four systems known to contain substellar donors, 50\% of them may have formed with a brown dwarf donor, and in progenitor systems with a similar mass ratio.

This does not fit with the observation of a ``brown dwarf desert'' \citep{duquennoy91,marcy00,grether06}, the lack of brown dwarfs observed in binaries with main sequence primaries and separations \textless\,3\,AU. The link between the ``brown dwarf desert" and CVs formed with brown dwarf secondaries is made by \cite{politano04}. Through population synthesis, \cite{politano04} find that the majority of progenitors of zero age CVs with brown dwarf secondaries have orbital separations and primary masses that coincide with this ``desert", explaining the dearth of CVs with substellar donors and periods below the period minimum.

The ``desert" is not completely arid, however, and a number of such systems do exist \citep{duchene13}. Taking all of this into consideration, we would expect to see significantly more post period-bounce systems than those formed directly with a brown dwarf donor, not the equal numbers found. This may be due to an observational bias against period-bounce CVs, but it is unclear what would cause this. It is therefore unlikely that PHL 1445 and SDSS 1433 are systems that formed with a brown dwarf donor, which opens up the possibility that both may actually be normal CVs, lying within the intrinsic scatter of the period minimum.

The current period minimum at $81.8 \pm 0.9$\,min (vertical dashed line in Figure~\ref{fig:m2vsp}) was determined by \cite{knigge11} through fitting a semi-empirical donor-based CV evolution track (red track in Figure~\ref{fig:m2vsp}) to the masses of a sample of CV donors. This sample of donors contains an intrinsic dispersion of $\sigma_{int}$ = 0.02 dex \citep{knigge11}, introducing an intrinsic scatter around the period minimum of equal value. This is equivalent to an intrinsic dispersion of $\sigma_{int}$ = 3.7\,min, significantly larger than the 0.9\,min error on the period minimum location. Approximately one third of this intrinsic scatter is due to the $\sim$20\% dispersion in white dwarf masses of the sample \citep{knigge06}. The majority of the remaining error can probably be attributed to a distribution in mass-loss rates, associated with residual magnetic braking below the CV period gap. This residual magnetic braking may explain why \cite{knigge11} require additional angular momentum loss below the period gap in order to produce a CV evolution track that is in agreement with the donor sample.

An independent measure of the intrinsic scatter can be obtained from the ``period spike'' analysis in \cite{gaensicke09}. The position of the period spike at $82.4 \pm 0.7$\,min is a good match to the period minimum from \cite{knigge11}, and \cite{gaensicke09} - assuming a Gaussian distribution - find a FWHM of 5.7\,min for this feature (see Figure~\ref{fig:m2vsp}). The intrinsic scatter on the period spike is therefore $\sigma_{int}$ = 2.4\,min. Using $\sigma_{int}$ of the period spike from \cite{gaensicke09} as the dispersion of systems around the period minimum, PHL 1445 and SDSS 1433 turn out to be 2.3$\sigma$ and 1.5$\sigma$ outliers, respectively.

In the sample of short period eclipsing CV systems in \cite{savoury11}, four systems have periods between 80 - 86\,mins, making them period spike systems according to \cite{gaensicke09}. Assuming PHL 1445 and SDSS 1433 are also systems near the period minimum brings the total to six. We must also assume here that no selection biases were involved with \cite{savoury11} choosing systems for model fitting. If CVs are distributed around a period minimum of 81.8\,min, with an intrinsic scatter of 2.4\,min, then the chances of finding these two outlying systems in such a small sample are approximately 6\%. This confirms the seemingly unlikely occurrence of finding two period minimum systems with periods as short as SDSS 1433 and PHL 1445, if the existing estimates for the position of the period minimum and $\sigma_{int}$ are correct.

It may be that the intrinsic scatter around the period minimum is underestimated, or that the position of the period minimum is incorrect. Being able to join the three substellar donor systems, PHL 1445, SDSS 1433 and SDSS 1035 with a single evolutionary track in Figure~\ref{fig:m2vsp} would provide evidence for the latter of these two possibilities, as this would suggest all three are period minimum systems that are of similar nature but just at different evolutionary stages. This would involve tweaking the parameters of the evolutionary model (e.g. angular momentum loss) and is beyond the scope of this paper, but the results would be of interest.

\section{Conclusions}
\label{sec:conclusions}

We have presented high-speed, three-colour photometry of the short-period eclipsing dwarf nova PHL 1445. Four eclipses were averaged to overcome the presence of flickering, making bright spot features visible and therefore enabling the determination of system parameters through eclipse model fitting. These system parameters include mass ratio $q$ = $0.087 \pm 0.006$, orbital inclination $i$ = $85.2 \pm 0.9^{\circ}$, primary mass $M_{w}$ = $0.73 \pm 0.03\,M_{\odot}$ and donor mass $M_{d}$ = $0.064 \pm 0.005\,M_{\odot}$, amongst others. The white dwarf temperature $T_{w}$ = $13200 \pm 700$\,K and photometric distance to the system $d$ = $220 \pm 50$\,pc were also found through multi-colour white dwarf flux fitting to model-atmosphere predictions.

We considered a number of possible reasons for PHL 1445 having an orbital period below the period minimum and determined their plausibility. PHL 1445's small proper motion does not make Galactic halo membership likely. Analysis of the donor's $r'$-band flux was used to rule out a significantly evolved donor, but one that is only slightly evolved ($X_{c}$ = 0.1) remains a possibility. Formation with a brown dwarf donor cannot be ruled out; although the brown dwarf would have to be older than 600\,Myrs at start of mass transfer, which requires a progenitor system with an extremely low mass ratio of $q$ = 0.025. Both PHL 1445 and SDSS 1433 - another CV with a substellar donor - lie below the period minimum for CVs, and their frequency may be evidence for error in the estimates for the intrinsic scatter and/or position of the period minimum.

\section{Acknowledgements}
\label{sec:acknowledgements}

MJM acknowledges the support of a UK Science and Technology Facilities Council (STFC) funded PhD. SPL, VSD and ULTRACAM are supported by STFC grant ST/J001589/1. TRM and EB are supported by STFC grant ST/L000733/1. SGP acknowledges financial support from FONDECYT in the form of grant number 3140585. The results presented in this paper are based on observations made with the William Herschel Telescope operated on the island of La Palma by the Isaac Newton Group in the Spanish Observatorio del Roque de Los Muchachos of the Instituto de Astrof\'isica de Canarias. This research has made use of NASA's Astrophysics Data System Bibliographic Services.

\bibliography{phl1445_references}

\end{document}